\documentstyle[11pt,aaspp4,psfig]{article}

\received{1999 Dec. 1}
\accepted{2000 Feb. 11}
\journalid{}{}
\articleid{}{}

\lefthead{Eracleous, Sambruna, \& Mushotzky}
\righthead{BLRGs With {\it RXTE}}

\begin{document}


\rightskip 0pt \pretolerance=100

\input psfig.tex

\font\hrm cmr12 at 26 truept
\font\lrm cmr12 at 18 truept
\font\smc  cmcsc10 at 20 truept

\leftline{\hskip -0.65truein \lrm 
P\hskip -1pt E\hskip -1 pt N\hskip -1pt N\hskip -1pt
{\hrm S}\hskip -1pt T\hskip -3pt A\hskip -3pt T\hskip -1pt E}
\vskip -8 truept
\leftline{\hskip -0.65truein \phantom{\lrm PENN}\hrulefill}
\vskip 0.15 truein
\hbox to 7truein{\vsize=1.truein
\newdimen\nameskip \nameskip=0.truein
\advance\nameskip by -\hoffset \hskip\nameskip
\vbox to 1. truein{\hsize=0.5truein
\vskip -0.15 truein 
\psfig{figure=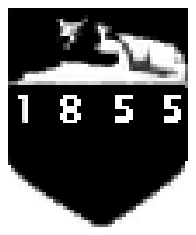,width=0.7in,rwidth=0.7in}
\vfill}
\vbox to 1. truein {\hsize=5truein \smc Astronomy and Astrophysics \hfill}}

\title{Hard X-Ray Spectra of Broad-Line Radio Galaxies from the 
Rossi X-Ray Timing Explorer}

\author{Michael Eracleous \& Rita Sambruna}
\affil{Department of Astronomy and Astrophysics, The Pennsylvania State
University, \\525 Davey Laboratory, University Park, PA 16802
\\e-mail: {\tt mce@astro.psu.edu, sambruna@astro.psu.edu}}

\and

\author{Richard F. Mushotzky}
\affil{NASA/GSFC, Code 662, Greenbelt, MD 20771
\\ e-mail: {\tt richard@xray-5.gsfc.nasa.gov}}

\medskip
\centerline
{To appear in {\it The Astrophysical Journal}, vol. 537, July 10, 2000}

\begin{abstract}
\rightskip 0pt \pretolerance=100 \noindent
We present the results of hard-X-ray observations of four
broad-line radio galaxies (BLRGs) with the {\it Rossi X-Ray Timing
Explorer} ({\it RXTE}). The original motivation behind the observations
was to search for systematic differences between the BLRGs and their
radio-quiet counterparts, the Seyfert galaxies. We do, indeed, find that
the Fe~K$\alpha$ lines and Compton ``reflection'' components, which are
hallmarks of the X-ray spectra of Seyferts galaxies, are weaker in BLRGs
by about a factor of 2. This observational result is in agreement with the
conclusions of other recent studies of these objects. We examine several
possible explanations for this systematic difference, including beaming of
the primary X-rays away from the accretion disk, a low iron abundance, a
small solid angle subtended by the disk to the primary X-ray source, and
dilution of the observed spectrum by beamed X-rays from the jet. We find
that a small solid angle subtended by the disk to the primary X-ray source
is a viable and appealing explanation, while all others suffer from
drawbacks. We interpret this as an indication of  a difference in the
inner accretion disk structure between Seyfert galaxies  and BLRGs, namely
that the inner accretion disks of BLRGs have the form of an ion-supported
torus or an advection-dominated accretion flow, which irradiates the
geometrically thin outer disk.
\end{abstract}

\clearpage

\section{Introduction}

An important issue in our study of active galactic nuclei (hereafter
AGNs) is the as yet unexplained difference between radio loud and
radio-quiet objects. All AGNs are thought to be powered by accretion of
matter onto a supermassive black hole, presumably via an equatorial
accretion disk. From a theoretical perspective, the accretion disk is an
essential ingredient for the formation of radio jets, although the exact
mechanism is not well known (see the review by Livio 1996). The
observational evidence\footnote{The most direct observational evidence
for the presence of accretion disks in AGNs takes the form of
Fe~K$\alpha$ lines with disk-like profiles in the X-ray spectra of
Seyfert galaxies (e.g., Tanaka et al. 1995; Nandra et al. 1997b) and
double-peaked H$\alpha$ lines in the optical spectra of BLRGs (Eracleous
\& Halpern 1994).} suggests that both radio-loud and radio-quiet AGNs
harbor accretion disks, but it is a mystery why well-collimated,
powerful, relativistic radio jets only exist in the former class of
object. The origin of the difference could lie in the nature of the host
galaxy. At low redshifts ($z<0.5$) radio-loud AGNs are found only in
elliptical galaxies, whereas radio-quiet AGNs can have either elliptical
or spiral hosts (Smith et al. 1986; Hutchings, Janson, \& Neff 1989;
V\'eron-Cetty \& Woltjer 1990; Dunlop et al. 1993; Bahcall et al. 1997;
Boyce et al. 1998). This observational trend has led to the suggestion
that the interstellar medium of the host galaxy may play an important
role in the propagation and collimation of the radio jets on large scales
(e.g., Blandford \& Levinson 1995; Fabian \& Rees 1995). Alternatively,
one may seek the fundamental cause of the difference between radio-loud
and radio-quiet AGNs in the properties of their accretion flows or the
properties of their central black holes. One possibility is that
radio-loud AGNs may harbor rapidly spinning black holes, whose energy is
extracted electromagnetically via the Blandford \& Znajek (1977)
mechanism and used to power the radio jets. Rapidly spinning black holes
could be associated with elliptical galaxies if both the black hole and
the host galaxy result from the merger of two parent galaxies, each with
its own nuclear black hole (see Wilson \& Colbert 1995).  Another
possibility is that the inner accretion disks of radio-quiet AGNs are
geometrically thin and optically thick throughout (Shakura \& Sunyaev
1973) while the inner accretion disks of  radio-loud AGNs are
ion-supported tori (Rees et al. 1982; known today as advection-dominated
accretion flows, or ADAFs, after the work of Narayan \& Yi 1994, 1995).
Because ADAFs are nearly spherical and parts of the flow are unbound,
they can lead to the formation of outflows (e.g., Blandford \& Begelman
1999). In either of the  above pictures, an additional mechanism may be
necessary to {\it collimate} the radio jets (see for example, Blandford
\& Payne 1982; Meier 1999).

If the difference between radio-loud and radio-quiet AGNs is related to 
differences in their central engines, it would lead to observable
differences in their respective X-ray spectra, in particular in the
properties (profiles and equivalent width) of the Fe~K$\alpha$ lines and
in the shape of the continuum. This is because the Fe~K$\alpha$ lines are
thought to result from fluorescence of the dense gas in the geometrically
thin and optically thick regions of the disk (e.g., George \& Fabian
1991; Matt et al 1992). Similarly, the continuum above 10~keV is thought
to include a significant contribution from X-ray photons from  the
``primary'' X-ray source, near the center of the disk, which undergo 
Compton scattering (``reflection'') in the same regions of the disk where
the  Fe~K$\alpha$ line is produced (e.g., Lightman \& White 1988; George
\& Fabian 1991; Matt, Perola, \& Piro 1991). It is, therefore, extremely
interesting that studies of the X-ray spectra of broad-line radio
galaxies (BLRGs) by Zdziarski et al. (1995) and Wo\'zniak et al. (1998)
found them to be systematically different from those of (radio-quiet)
Seyfert  galaxies. In particular, these authors found that the signature
of Compton reflection, which is very prominent in the spectra of Seyfert
galaxies above 10~keV (Pounds et al. 1989; Nandra \& Pounds 1994), is
weak of absent in the spectra of BLRGs. Moreover, Wo\'zniak et al. (1997)
found that the Fe~K$\alpha$ lines of BLRGs are  narrower and weaker than
those of Seyfert galaxies.

Motivated by the above theoretical considerations and observational
results, we have undertaken a systematic study of the X-ray spectra of
radio-loud AGNs with {\it ASCA} and {\it RXTE} in order to characterize
their properties. Our  main goal is to compare their spectroscopic
properties with those of Seyfert galaxies and test the above ideas for
the origin of the difference between the two classes. In our re-analysis
of archival {\it ASCA} spectra of BLRGs and radio-loud quasars (Sambruna,
Eracleous, \& Mushotzky 1999)  we found that the Fe~K$\alpha$ lines of
some objects are indeed weaker and narrower than those of Seyferts, in
agreement with the findings of Wo\'zniak et al. (1997). In other 
objects, however, the uncertainties are large enough that we cannot reach
firm conclusions, thus we have not been able to confirm this result in
general. In this paper we present the results of new observations of four
BLRGs with {\it RXTE}, aimed at measuring the shape of their hard X-ray
continuum and the equivalent width of their Fe~K$\alpha$ lines. As such,
these observations complement our study of the {\it ASCA} spectra of
these objects. In \S2 we describe the observations and data screening. In
\S3 we present and discuss the light curves and in \S4 we compare the
observed spectra with models. In \S5 we discuss the implications of the
results, while in \S6 we summarize our conclusions.  Throughout this
paper we assume a Hubble constant of $H_0=50~{\rm km~s^{-1}~Mpc^{-1}}$ and
a deceleration parameter of $q_0=0.5$.


\begin{deluxetable}{lcrcccc}
\tablenum{1}
\tablewidth{6.5in}
\tablecolumns{6}
\tablecaption{Target Objects, Basic Properties, and Observation Log}
\tablehead{
 & & & \multicolumn{2}{c}{$N_{\rm H}$ (cm$^{-2}$)} & \nl
 & & & \multicolumn{2}{l}{\hrulefill} & \colhead{Observation} & Duration \nl
\colhead{Object}  &  
\colhead{$z$}  &
\colhead{$i$\tablenotemark{\;a}}  & 
\colhead{Galactic\tablenotemark{\;b}}  & 
\colhead{{\it ASCA}\tablenotemark{\;c}} &
\colhead{Start Time (UT)} &
\colhead{(hours)}
}
\startdata
3C 111    & 0.048 & $37^{\circ}>i>24^{\circ}$           & $1.2\times 10^{22}$ & $9.63\times 10^{21}$ & 1997/3/22 01:23 & 62 \nl
3C 120    & 0.033 & $14^{\circ}>i>1^{\circ}\phantom{4}$ & $1.2\times 10^{21}$ & $1.65\times 10^{21}$ & 1998/2/13 04:53 & 58 \nl
Pictor A  & 0.035 & $i>24^{\circ}$                      & $4.2\times 10^{20}$ & $8.30\times 10^{20}$ & 1997/5/08 02:21 & 82 \nl
3C 382    & 0.057 & $i>15^{\circ}$                      & $6.7\times 10^{20}$ & $6.70\times 10^{20}$ & 1997/3/28 23:26 & 47 \nl
\tablenotetext{a\;} {The inclination angle of the radio jet (see Eracleous \&
                    Halpern 1998, and references therein).}
\tablenotetext{b\;} {The Galactic equivalent H{\sc\, i} column density.
                    {\it References}: 
                    3C~111: Bania, Marscher, \& Barvainis (1991);
                    3C~120: Elvis et al. (1989);
                    Pictor~A: Heiles \& Cleary (1979);
                    3C 382: Murphy et al. (1996).
                    }
\tablenotetext{c\;} {The Galactic equivalent H{\sc\, i} column density 
                    as measured by the {\it ASCA} SIS (taken from 
                    Sambruna et al. 1999).}
\enddata
\end{deluxetable}


\begin{deluxetable}{lcccccc}
\tablenum{2}
\tablewidth{6.5in}
\tablecolumns{7}
\tablecaption{Exposure Times and Count Rates}
\tablehead{
 & \multicolumn{3}{c}{PCA (2.5--30 keV)} & 
\multicolumn{3}{c}{HEXTE cluster 0 (20-100 keV)}\nl
 & \multicolumn{3}{c}{\hrulefill} & \multicolumn{3}{c}{\hrulefill}\nl
\colhead{} &
\colhead{Exp.} &
\colhead{Source} &
\colhead{Backg.} &
\colhead{Exp.} &
\colhead{Source} &
\colhead{Backg.} \nl
\colhead{} &
\colhead{Time} &
\colhead{Count Rate} &
\colhead{Count Rate} &
\colhead{Time} &
\colhead{Count Rate} &
\colhead{Count Rate} \nl
\colhead{Object} &
\colhead{(s)} &
\colhead{(s$^{-1}$ PCU$^{-1}$)} &
\colhead{(s$^{-1}$ PCU$^{-1}$)} &
\colhead{(s)} &
\colhead{(s$^{-1}$)} &
\colhead{(s$^{-1}$)} 
}
\startdata
3C 111    & 33,440 & 22.9 & 39.1 & 12,748 & 7.5 & 158.4 \nl
3C 120    & 55,904\tablenotemark{\;a} & 25.2 & 39.9 & 17,572 & 3.7 & 187.5 \nl
Pictor A  & 30,240\tablenotemark{\;a} & 9.6 & 38.7 & 12,461 & 9.4 & 166.9 \nl
3C 382    & 28,192 & 13.9 & 22.0 & 10,606 & 3.0 & 114.9 \nl
\tablenotetext{a\;} {The PCA exposure time refers to PCUs 0, 1, and 2.
                     PCUs 3 and 4 were off during part of the observation.}
\enddata
\end{deluxetable}


\section{Targets, Observations, and Data Screening}

Our targets were selected to be among the X-ray brightest BLRGs, since
the background in the {\it RXTE} instruments is rather high. They are
listed in Table~1 along with their basic properties, namely the redshift,
the inferred inclination of the radio jet (see Eracleous \& Halpern 1998a
and references therein), and  two different estimates of the column
density in the Galactic interstellar medium. The 2--10~keV fluxes are
about a few $\times 10^{-11}~{\rm erg~s^{-1}~cm^{-2}}$,  which makes the
targets readily detectable by the  {\it RXTE} instruments. The
observations were carried out with the Proportional Counter Array (PCA;
Jahoda et al. 1996) and the High-Energy X-Ray Timing Experiment (HEXTE;
Rothschild et al. 1998) on {\it RXTE}. Three out of the four objects in
our collection (3C~111, Pictor~A, and 3C~382) were observed in the spring
of 1997 as part of our own guest-observer programs. The fourth object
(3C~120) was observed in 1998 February as part of a different program and
the data were made public immediately. Although a fair number of
observations of 3C~120 exist in the {\it RXTE} archive, only the above
observation was a long, continuous observation suitable for our purposes.
All other observations  were intended to monitor the variability of this
object; they consist of  short snapshots spanning a long temporal
baseline. Similarly, the only  observations of 3C~390.3 available in the
archive are also monitoring  observations of this type. Therefore we have
not included 3C~390.3 in our  sample. A log of the observations is
included in Table~2. All of the observations fall within PCA gain epoch~3.

The PCA data were screened to exclude events recorded when the pointing
offset was greater than 0\fdg02, when the Earth elevation angle was less
than 10\arcdeg, or when the electron rate was greater than 0.1. Events
recorded within 30 minutes of passage though the South-Atlantic Anomaly
were also excluded. After screening, time-averaged spectra were extracted
by accumulating events recorded in the top layer of each  Proportional
Counter Unit (PCU) since these include about 90\% of the source photons
and only  about 50\% of the internal instrument background. In the case
of 3C~111 and 3C~382 we extracted light curves and spectra from all five
PCUs.  In the case of 3C~120 and Pictor~A, PCUs 3 and 4 were turned off
part of the time, therefore the spectra from these two PCUs and PCUs 0,
1, and 2 were accumulated separately and then added together. Light
curves corresponding to the energy range 3.6--11.6~keV (PCA channels
6--27), were also extracted for all objects and rectified over the time
intervals when one or more of the PCUs were not turned on.  The PCA
background spectrum and light curve were determined using the
\verb+L7_240+ model developed at the {\it RXTE} Guest-Observer Facility
(GOF) and implemented by the program \verb+pcabackest v.2.1b+.  This
model is appropriate for ``faint'' sources, i.e., those producing count
rates less than 40~s$^{-1}$~PCU$^{-1}$.  All of the above tasks were
carried out using the \verb+FTOOLS v.4.2+ software package and with the
help of the \verb+rex+ script provided by the {\it RXTE}~GOF, which also
produces response matrices and effective area curves appropriate for the
time of the observation. The net exposure times after data screening, as
well as the total source and background count rates, are given in
Table~2. 

The HEXTE data were also screened to exclude events recorded when the
pointing  offset was greater than 0\fdg02, when the Earth elevation angle
was less than 10\arcdeg. The background in the two HEXTE clusters is
measured during each observation by rocking the instrument slowly on and
off source. Therefore, source and background photons are included in the
same data set and are separated into source and background spectra
according to the time they were recorded.  After screening we extracted
source and background spectra from each of the two HEXTE clusters and
corrected them for the dead-time effect, which can be significant because
of the high background rate. As with the PCA data, the HEXTE data were
screened and reduced using the \verb+FTOOLS v.4.2+ software package. The
exposure times and count rates in HEXTE/cluster~0 are given in Table~2
for reference. In Figure~1 we show the on-source and background spectra
from the PCA and HEXTE/cluster~0, using 3C~111 as an example. This figure
illustrates that the background makes the dominant contribution to the
HEXTE count rate. In the case of the  PCA, the background contributes
approximately the same count rate as the source, and its relative
contribution increases with energy. At energies above 30~keV the
background dominates the count rate in the PCA.

\begin{figure}[t]
\centerline{\psfig{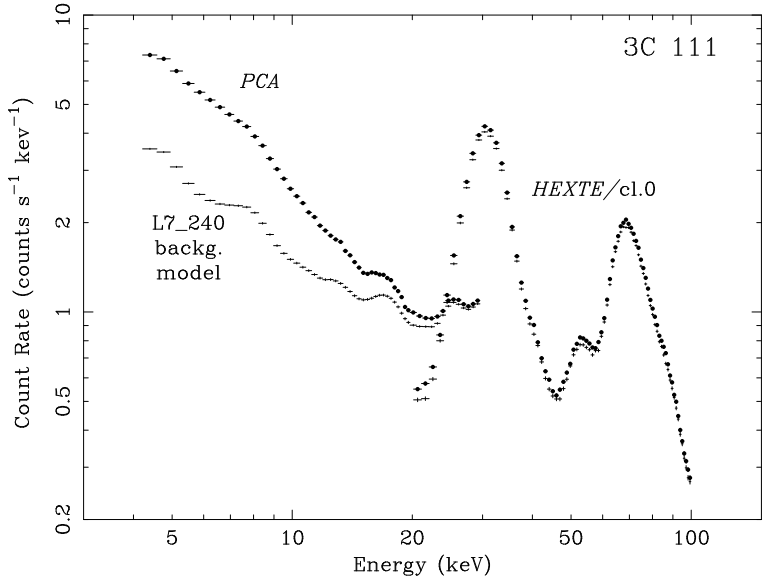}}
\caption{\footnotesize 
The PCA and HEXTE count rate spectra of 3C~111 as examples of typical
data obtained from the {\it RXTE} instruments. The on-source spectrum, shown as
large dots, is compared to the background spectrum. In the case of the PCA, the
background spectrum was computed using the  {\tt L7\_240} model, while in the
case of the HEXTE the background spectrum was measured during the observation
by rocking the instrument on and off source. Notice that the contribution of 
the background to the total count rate in the PCA increases with energy so
that above 30~keV the observed count rate is dominated by the background.
In the case of HEXTE, the background dominates the total count rate at all
energies.}
\end{figure}

\section{Light Curves and Time Variability}

In Figure~2 we show the light curves of the four targets. For each
object we plot the net count rate (after background subtraction) as well
as the the background count rate {\it vs} time, for reference. The {\it
mean} background count rate is generally comparable to the net source
count rate, although the background level varies dramatically over the
course of an observation. A visual inspection of the object light curves
shows no obviously significant variability. To quantify this result we
searched the light curves for variability using two complementary methods:

\begin{figure}
\centerline{\psfig{file=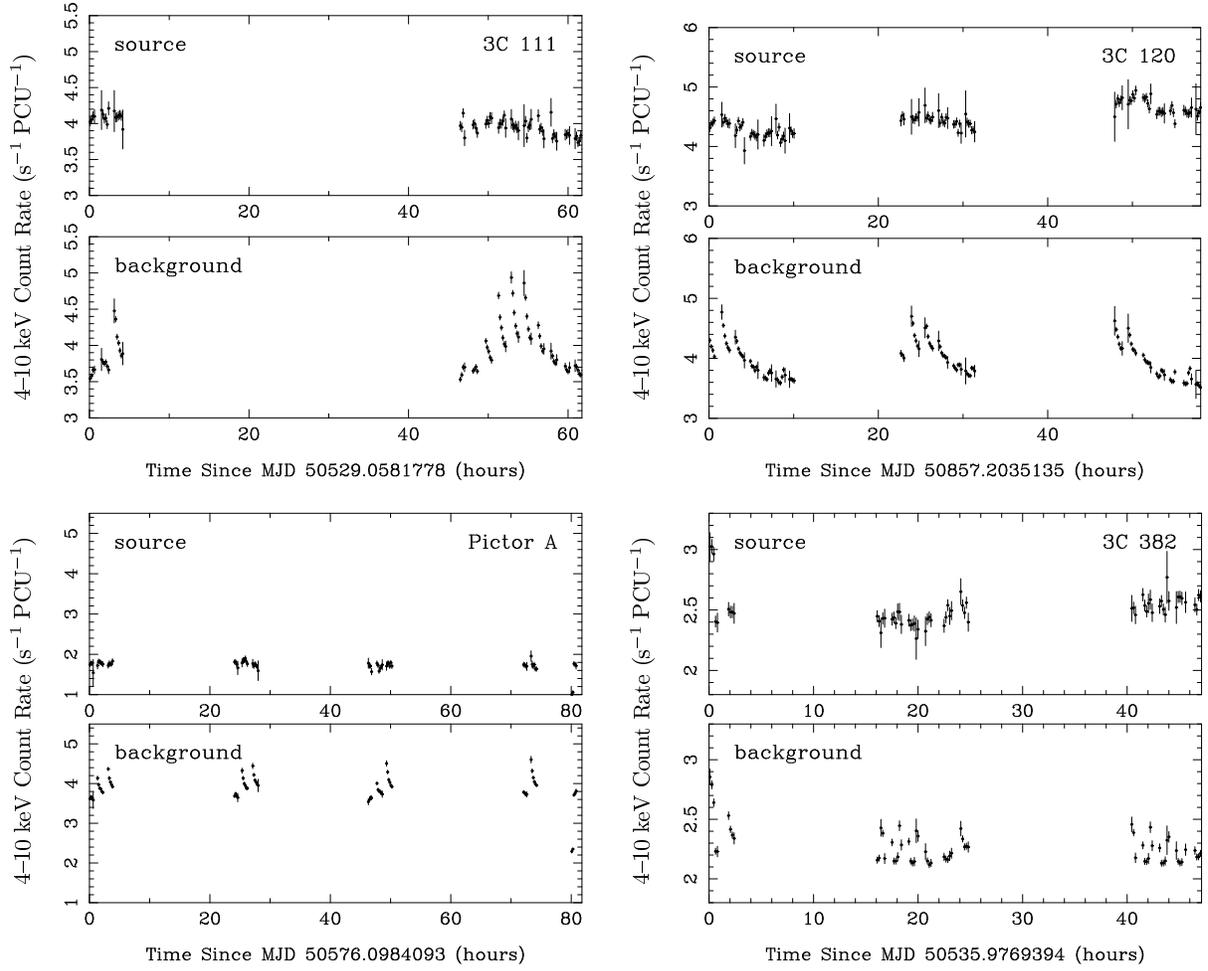,width=7.5in,rheight=5.5in,angle=90}}
\caption{\footnotesize
Light curves showing the 4--10~keV count rate per PCU of the
target BLRGs as a function of time over the course of the {\it RXTE}
observation. For each object we plot the net source light curve and the
background light curve computed from the {\tt L7\_240} model for reference}
\end{figure}

\begin{enumerate}
\rightskip 0pt \pretolerance=100 

\item
We compared the variance (i.e., the r.m.s. dispersion about the mean)
with the average uncertainty of points in the light curve. This method
is sensitive to fluctuations in the count rate that exceed the noise
level on time scales comparable to the width of the bins in the light
curve (the bin width in the light curves of Figure~2 is 640~s). We
found no large fluctuations in any of the light curves, save for two
instances of poor background subtraction (in Pictor~A and 3C~382). We
have also computed the ``excess variance'', $\sigma^2_{\rm rms}$,
following Nandra et al. (1997a), with the following results:
$(3.1\pm0.1)\times 10^{-4}~{\rm s}^{-2}$ for 3C~111,
$(1.23\pm0.01)\times 10^{-3}~{\rm s}^{-2}$ for 3C~120,
$(-8.3\pm0.6)\times 10^{-4}~{\rm s}^{-2}$ for Pictor~A, and
$(6.1\pm0.4)\times 10^{-4}~{\rm s}^{-2}$ for 3C~382. Since their
luminosities are $L_{\rm x}$(2--10~keV)$\sim 10^{44}~{\rm erg~s^{-1}}$,
these objects fall on the extrapolation of the $\sigma^2_{\rm
rms}$--$L_{\rm x}$ trend for Seyfert galaxies, found by Nandra et
al. (1997a). It is noteworthy, however, that the excess variance is
small enough at this luminosity that it is comparable to that of LINERs
and other very low-luminosity AGNs (Ptak et al. 1998).

\item
We fitted each light curve with a polynomial to find the lowest order
that gives an acceptable fit. This method detects small, relatively
slow variations in the light curve on time scales somewhat shorter
than the length of the observations. We found that 3C~111, Pictor~A,
and 3C~382 show small secular variations on the order of a few percent
relative to the mean while 3C~120 shows relatively slow variations
with excursions of $\pm6$\% from the mean. For reference, we note that
previous observations of 3C~120 with {\it ASCA} have shown it to be
variable at the 20\% level (Grandi et al. 1997).

\end{enumerate}

\noindent
The above results are not surprising since BLRGs are generally known
not to be highly variable on short time scales, although they can vary
substantially on time scales of several days (see for example the soft
X-ray light curve of 3C~390.3 presented by Leighly \& O'Brien 1997).
We will return to the issue of variability in our later discussion of
our overall findings.

\section{Model Fits to the Observed Spectra}

\subsection{The Continuum Shape}

The shape of the continuum was determined by fitting models to the
observed spectra with the help of the \verb+XSPEC v.10.0+ software
package (Arnaud 1996).  We used PCA response matrices and effective
area curves created specifically  for the individual observations by
the program \verb+pcarsp v.2.37+, taking into account the evolution of
the detector properties. All the spectra from individual PCUs were
added together and the individual response matrices and effective area
were combined accordingly. In fitting the HEXTE spectra we used the
response matrices and effective area curves created on 1997 March 20.
The spectra from the two HEXTE clusters were not combined. All spectra
were rebinned so that each bin contained enough counts for the $\chi^2$
test to be valid. The PCA spectra were truncated at low energies at
4~keV and at high energies at either 20 or 30~keV depending  on the
signal-to-noise ratio. In the case of the HEXTE spectra we retained the
energy channels between 20 and 100~keV. 

We compared the spectra with models consisting of a continuum component
and a Gaussian line, modified by interstellar photoelectric absorption.
We adopted the absorbing column densities measured by {\it ASCA}
(listed in Table~1), which we held fixed throughout the fitting
process. These column densities are comparable to or greater than the
Galactic column densities inferred from \ion{H}{1}~21~cm observations
(see Table~1). The photoelectric absorption cross-sections used were
those of Morrison \& McCammon (1983). We tried three different models
for the continuum shape: a simple power law, a broken power law, and a
power law plus its Compton reflection from dense, neutral matter. The
Compton reflection model is meant to describe the effects of Compton
scattering of photons from an X-ray source associated with the inner
parts of the accretion disk in the gas that makes up the disk proper
(e.g., Lightman \& White 1988; George \& Fabian 1991). The spectrum of
reflected X-rays was computed as a function of disk inclination using
the transfer functions of  Magdziarz \& Zdziarski (1995)\footnote{The
Compton reflection calculation is carried out by the \verb+pexrav+
model routine in \verb+XSPEC+}. The free parameters of this model, in
addition to the spectral index of the primary power law and the
inclination angle of the disk, are the folding (i.e., upper cut-off)
energy of the primary X-ray spectrum, the solid angle subtended by the
disk to the central X-ray source, and the abundances of iron and other
heavy elements. In our fits the inclination angles were constrained to
lie within the limits inferred from the radio properties of each object
(see Table~1), and the folding energy of the primary X-ray spectrum was
constrained to be greater than 100~keV since all objects are detected
by HEXTE up to that energy. The broken power law model is effectively a
parameterization of  the Compton reflection model: we included it in
our suite of continuum models because it serves as an additional
verification of the possible departure of the continuum shape from a
simple power law. Although the broken power law and Compton reflection
models have different shapes in detail, the difference is not
discernible at the  signal-to-noise ratio of the available HEXTE
spectra.

The results of fitting the above models to the observed spectra are
summarized in Table~3. The fits yield 2--10~keV unabsorbed fluxes for
the target objects in the range
2--6$\times 10^{-11}$~erg~cm$^{-2}$~s$^{-1}$ and corresponding
luminosities in the range 1--5$\times 10^{44}$~erg~s$^{-1}$, as listed
in Table~3. Spectra with models superposed, are shown in Figures~3 and
4. We find that the spectra of two objects, 3C~111 and Pictor~A, are
described quite well by a simple power law throughout the observed
energy range (4--100~keV), as shown in Figure~3. A broken power law
model produces a slightly better fit but the improvement is not
statistically significant in view of the additional free parameters
(the F-test gives chance improvement probabilities of 0.35 and 0.29
respectively). Similarly, the Compton reflection model does not produce
a significantly improved fit either.\footnote{In fact, the Compton
reflection fit results in a higher value of $\chi^2$ than the broken
power law, even though it has more free parameters.} In the case of the
other two objects, 3C~120 and 3C~382, we find that a simple power law
does not provide an adequate description of the continuum shape (see
Figure~3); a broken power law or a Compton reflection model is required
by the data (the F-test gives chance improvement probabilities for the
Compton reflection model of $5\times 10^{-5}$ and $5\times 10^{-3}$,
respectively). Fits of power-law plus Compton reflection models to the
spectra of these two objects are shown in Figure~4.  Finally, we note
that all of the BLRGs in our collection were also observed with {\it
SAX} and detected at high energies up to 50~keV (Padovani et al. 1999;
Grandi et al. 1999a,b). The {\it SAX} spectra yield very similar results
to what we obtain here, namely very similar spectral indices and
Compton reflection strengths.

\clearpage 

\begin{deluxetable}{lllll}
\tablenum{3}
\tablewidth{6.5in}
\tablecolumns{5}
\tablecaption{Results of Model Fits\tablenotemark{\;a}}
\tablehead{
\colhead{Model and Parameters} &
\colhead{3C 111} &
\colhead{3C 120} &
\colhead{Pictor A} &
\colhead{3C 382}
}
\baselineskip 23pt
\startdata
\cutinhead{\sc Continuum Models}
\sidehead{\it Simple Power Law}
Spectral Index, $\Gamma$               & $1.76\pm0.01$ & $1.82\pm0.01$ & $1.80\pm0.03$ & $1.81\pm0.02$ \nl
Flux density at 1 keV ($\mu$Jy)\tablenotemark{\;b} & 8.9$\pm0.2$ & 10.6$\pm0.2$ & 3.9$\pm0.3$ & 5.6$\pm0.2$ \nl
Total and Reduced $\chi^2$             & 52.30, 0.84   & 125.5, 2.02   & 38.51, 0.79   & 68.83, 1.41 \nl
\sidehead{\it Broken Power Law}
Low-Energy Spectral Index, $\Gamma_1$  & $1.77\pm0.02$ & $1.87^{+0.05}_{-0.02}$ & $1.77^{+0.04}_{-0.05}$ & $1.86^{+0.03}_{-0.04}$ \nl
High-Energy Spectral Index, $\Gamma_2$ & $1.7\pm0.1$   & $1.7\pm0.1$            & $2\pm1$                & $1.5^{+0.1}_{-0.2}$\nl
Break Energy (keV)                     & $11\pm5$      & $9\pm1$                & $11\pm3$               & $10\pm2$ \nl
Flux density at 1 keV ($\mu$Jy)\tablenotemark{\;b} & 9.0$\pm0.2$ & 11.5$\pm0.2$ & 3.8$\pm0.3$ & 6.1$\pm0.2$ \nl
Total and Reduced $\chi^2$             & 47.79, 0.80 & 58.55, 0.98 & 33.53, 0.71 & 40.81, 0.73 \nl
\sidehead{\it Power Law + Compton Reflection\tablenotemark{\;c}}
Spectral Index, $\Gamma$               & $1.78^{+0.03}_{-0.05}$ & $1.90\pm0.03$ & $1.72^{+0.04}_{-0.03}$ & $1.92^{+0.07}_{-0.09}$ \nl 
Folding Energy (keV)\tablenotemark{\;d} & 1000 ($>100$) & 15,700 ($>300$) & $>100$ & 300 ($>100$) \nl 
Reflector Solid Angle, $\Omega/2\pi$   & $0.1^{+0.4}_{-0.1}$ & $0.4^{+0.4}_{-0.1}$ & $<0.3$ & $0.5^{+0.8}_{-0.2}$ \nl 
Inclination Angle, $i$\tablenotemark{\;e} & 24\arcdeg & 10\arcdeg & 60\arcdeg & 16\arcdeg \nl 
Flux density at 1 keV ($\mu$Jy)\tablenotemark{\;b} & 9.1$\pm0.2$ & 11.8$\pm0.2$ & 3.3$\pm0.2$ & 6.5$\pm0.2$ \nl
Total and Reduced $\chi^2$             & 47.75, 0.81 & 64.37, 1.09 & 35.61, 0.76 & 40.46, 0.88 \nl
\sidehead{}
2--10 keV Flux (${\rm erg~s^{-1}~cm^{-2}}$)\tablenotemark{\;f} & $5.6\times 10^{-11}$ & $6.1\times 10^{-11}$ & $2.3\times 10^{-11}$ & $3.3\times 10^{-11}$ \nl
2--10 keV Luminosity (${\rm erg~s^{-1}}$)\tablenotemark{\;f} & $5.5\times 10^{44}$ & $2.9\times 10^{44}$ & $1.2\times 10^{44}$ & $4.6\times 10^{44}$ \nl
\cutinhead{\sc Gaussian Emission-Line Model}
Rest Energy Dispersion, $\sigma_{\rm rest}$ (keV) & $<0.9$ & $<1.2$ & $<1.5$ & $<1.2$ \nl
FWHM (km s$^{-1}$)                        & $<44,000$ & $<55,000$ & $<70,000$ & $<56,000$ \nl
Line Flux ($10^{-5}~{\rm photons~s^{-1}~cm^{-2}}$)& $3.2^{+1.1}_{-0.5}$ & $4.9^{+1.6}_{-1.1}$ & $1.8\pm0.8$ & $3.1\pm0.9$ \nl
Rest Equivalent Width (eV)                & $60^{+20}_{-10}$ & $90^{+30}_{-20}$    & $80\pm40$ & $90\pm30$ \nl
\tablenotetext{a\;} 
{All error bars and limits correspond to the 90\% confidence level.}
\tablenotetext{b\;}
{The flux density at 1 keV was derived from the normalization of the PCA 
spectrum. It is related to the monochromatic photon flux at 1keV
via $f_{\nu}=h\,E\,N_{\rm E}$, where $h$ is Planck's constant. If 
$N_{\rm E}$ is measured in ${\rm photons~s^{-1}~cm^{-2}~keV^{-1}}$,
then $f_{\nu}(1\;{\rm keV})=663\;N_{\rm E}~\mu{\rm Jy}$.}
\tablenotetext{c\;}
{The results reported here were obtained by keeping the
abundances of iron and other heavy elements fixed at the solar 
value.}
\tablenotetext{d\;}
{Best-fitting value and lower limit to the folding energy.}
\tablenotetext{e\;}
{The inclination angles were restricted to lie in the range given in 
Table~1. In the case of Pictor~A and 3C~382, 
we assumed conservatively that $i<60\arcdeg$.}
\tablenotetext{f\;}
{The observed 2--10~keV fluxes and luminosities turn out to be the same 
for all models. The luminosities reported here have been corrected for
absorption.}
\enddata
\end{deluxetable}

\clearpage

\begin{figure}
\centerline{\psfig{file=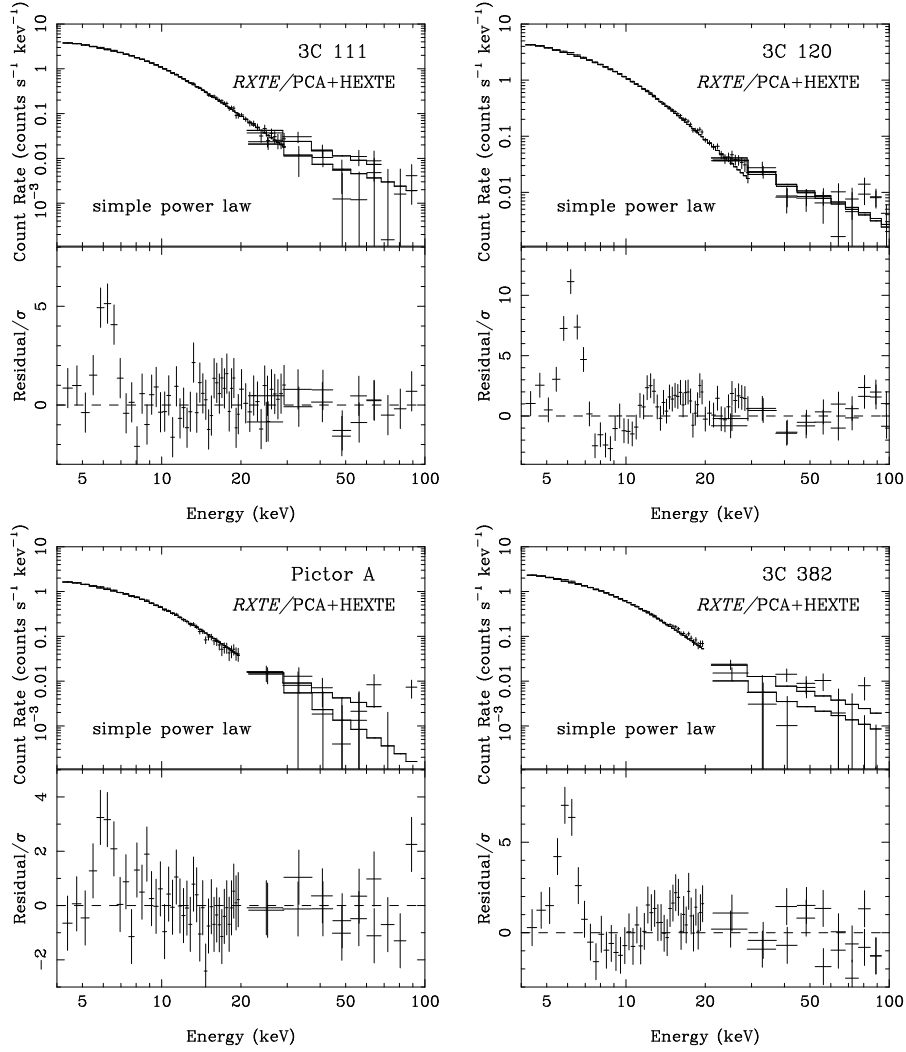,rheight=6.5in,height=8in}}
\caption{\footnotesize
The 4--100~keV spectra of the target objects with the best
fitting simple power-law continuum model superposed. The model continuum is
also modified by interstellar photoelectric absorption. The PCA
spectra cover the range 4--30~keV or 4--20~keV (finely binned) while
the HEXTE spectra cover the range 20--100~keV (coarsely binned). The
lower panel in each set shows the residual spectrum after subtraction
of the continuum model, in which the Fe~K$\alpha$ line is
obvious. The residual count rate at each point has been scaled by the error
bar. In the case of 3C~111 and Pictor~A, the simple power-law model
provides a good description of the continuum. In the case of 3C~120
and 3C~382 however, this model does not provide an adequate description
of the continuum: there are substantial negative residuals at energies
just below 10~keV and substantial positive residuals at energies between
10 and 20~keV.}
\end{figure}

\clearpage

\begin{figure}[t]
\centerline{\psfig{file=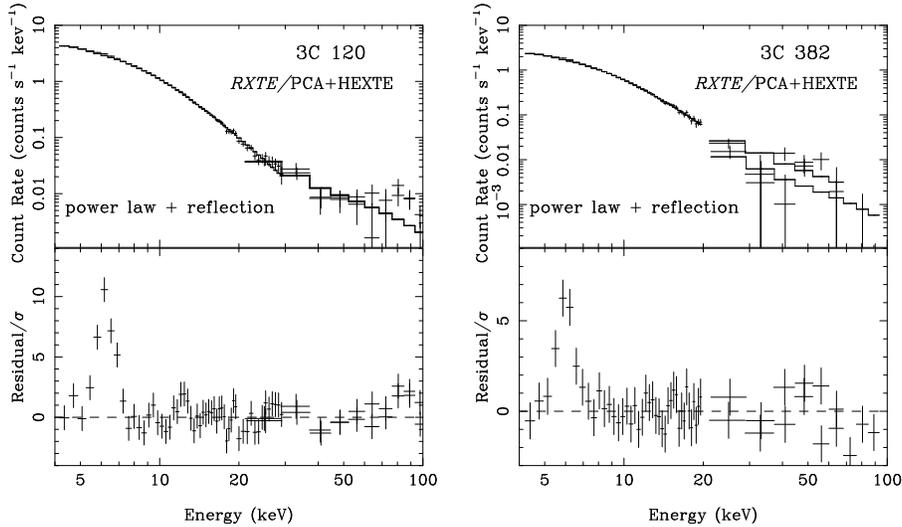,rheight=3.5in,height=8in}}
\caption{\footnotesize
The 4--100~keV spectra of 3C~120 and 3C~382 with the best fitting
power-law plus Compton reflection model superposed. The model
continuum is also modified by interstellar photoelectric
absorption. The PCA spectra cover the range 4--30~keV or 4--20~keV
(finely binned) while the HEXTE spectra cover the range 20--100~keV
(coarsely binned). The lower panel in each set shows the residual
spectrum after subtraction of the continuum model, in which
the Fe~K$\alpha$ line is obvious. The residual count rate at each
point has been scaled by the error bar. This model provide a good
description of the continuum in these two objects, unlike the simple
power-law model shown in Figure~3.}
\end{figure}

\begin{figure}
\centerline{\psfig{file=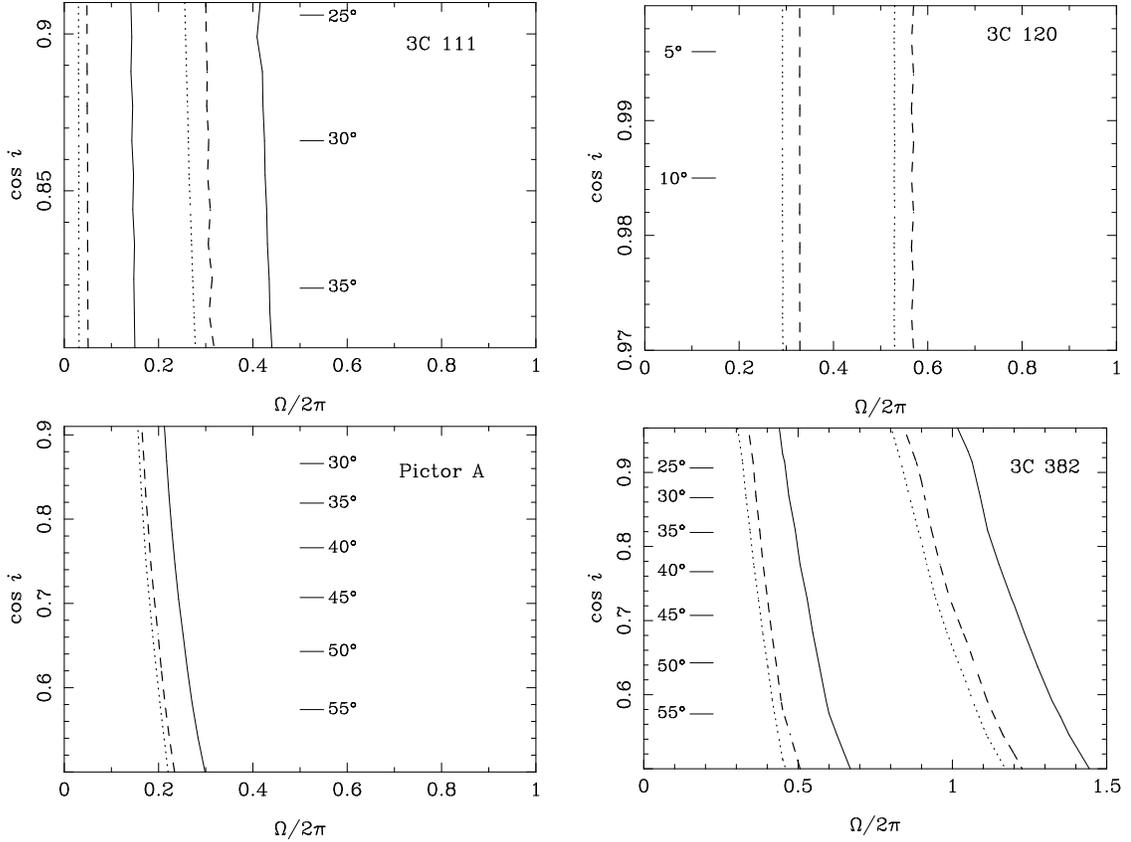,width=7.5in,rheight=5.5in,angle=-90}}
\caption{90\% confidence contours model in the $\cos i$--$\Omega/2\pi$
plane for the Compton reflection. The range of $\cos i$ plotted on the
vertical  axis corresponds to the allowed values listed in Table~1. In the
cases of Pictor~A and 3C~382 we have assumed conservatively that
$i<60\arcdeg$. The region of acceptable fits is bounded by two lines of the
same style except for Pictor~A, where the lines represent upper limits on
$\Omega/2\pi$. Different line styles represent different assumed values of
the folding energy of the primary power-law spectrum as follows: solid: 
$E_{\rm fold}=100$~keV, dashed: $E_{\rm fold}=300$~keV, and dotted:
$E_{\rm fold}=600$~keV. In the case of 3C~120 the data imply that
$E_{\rm fold}>300$~keV, therefore the $E_{\rm fold}=100$~keV contour is
not plotted.}
\end{figure}

\begin{figure}[t]
\centerline{\psfig{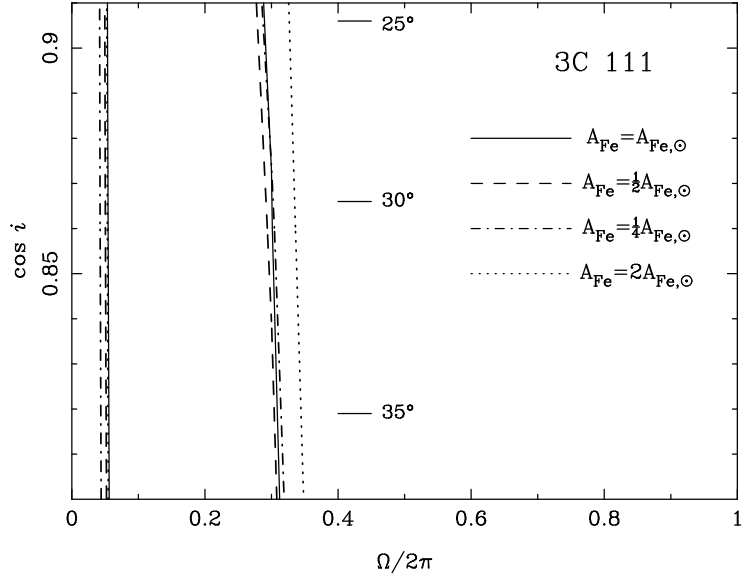}}
\caption{An illustration of the negligible effect of the Fe abundance on
the confidence contours in the $\cos i$--$\Omega/2\pi$ plane, using 3C~111
as an example. The solid lines correspond to a Solar iron abundance, the
dashed lines to half Solar, the dash-dot lines to a quarter of Solar, and
the dotted lines to twice Solar. The results are nearly identical,
independently of the Fe abundance.}
\end{figure}

We have explored the multidimensional space defined by the free
parameters of the Compton reflection model to find the range of
acceptable parameter values. We found that the folding energy of the
primary X-ray spectrum is  not constrained very well by the observed
spectra. Thus our initial physical restriction of $E_{\rm
fold}>100$~keV is the only constraint on the folding energy of the
primary power-law spectrum. Only in the case of 3C~120 (the brightest
object, with the longest exposure time) are we able to derive a
somewhat better constraint from the data of $E_{\rm fold}>300$~keV. We
are particularly interested in the solid angle subtended by the
reflector to the primary X-ray source, which is a diagnostic of the
geometry and structure of the accretion flow. This  parameter
determines the {\it absolute} strength of the reflected component. In
practice, however, it is not straightforward to constrain the reflector
solid angle because the inclination angle of the disk and the iron
abundance affect the {\it observed} strength of this component, through
projection and photoelectric absorption above the Fe K edge at 7.1~keV
(see, for example, George \& Fabian 1991; Reynolds, Fabian, \& Inoue
1995), respectively. Moreover, the folding energy of the primary X-ray
spectrum also affects the determination of the reflector solid angle
because it controls the number of energetic photons available for
Compton downscattering. We have, therefore, examined the effect  of
each of the above parameters on our derived values of the reflector
solid angle by searching the inclination angle - solid angle ($\cos
i$--$\Omega/2\pi$) plane for regions with acceptable fits for different
assumed values of the folding energy and iron abundance. In other
words, we have taken 2-dimensional slices of the 4-dimensional
parameter space defined by the solid angle, inclination angle, folding
energy, and iron abundance. In Figure~5 we show  the 90\% confidence
contours in the $\cos i$--$\Omega/2\pi$ plane  for a Solar iron
abundance and three different assumed values of the folding energy 
($E_{\rm fold}=100$, 300, and 600~keV). The same results are also
summarized in Table~3. The strength of the Compton reflection
component,  as parameterized by $\Omega/2\pi$ is systematically lower
than  what is found  in Seyferts observed with either {\it Ginga} or
{\it RXTE}. In particular, we find that $\Omega/2\pi\lesssim 0.5$,
while in Seyferts $\Omega/2\pi\approx  0.8$, with a dispersion of 0.2
(Nandra \& Pounds 1994; see also Weaver, Krolik, \& Pier  1998 and Lee
et al. 1998, 1999 for results from {\it RXTE} observations). 3C~382 is
the only case where the measured  value of $\Omega/2\pi$ is consistent,
within uncertainties, with what is found in Seyferts. Pictor~A
represents the opposite extreme where only an upper limit of
$\Omega/2\pi<0.3$ can be obtained. In the case of 3C~111, the 
detection of a Compton reflection hump should be regarded as only
marginal since (a) it is only detected at the 90\% confidence level and
is consistent with zero at the 99\% confidence level for $E_{\rm
fold}=300,\,600$~keV, and  (b) the probability of chance improvement of
using the Compton-reflection model compared to the power-law model is
0.39 according to the F-test. In the case of 3C~111 and 3C~382 our
results are consistent with what Nandra \& Pounds (1994) and Wo\'zniak
et al. (1998) report based on {\it Ginga} observations.  The iron
abundance does not affect the above conclusions at all. We find that 
the confidence contours in the $\cos i$--$\Omega/2\pi$ plane are
nearly  identical for iron abundances ranging from one quarter to twice
the Solar value. We illustrate this result in Figure~6 using 3C~111 as
an example. This issue is relevant to the strength of the Fe~K$\alpha$
lines whose measurements we report in the next section.

\begin{figure}[t]
\vskip -0.5truein
\centerline{\psfig{file=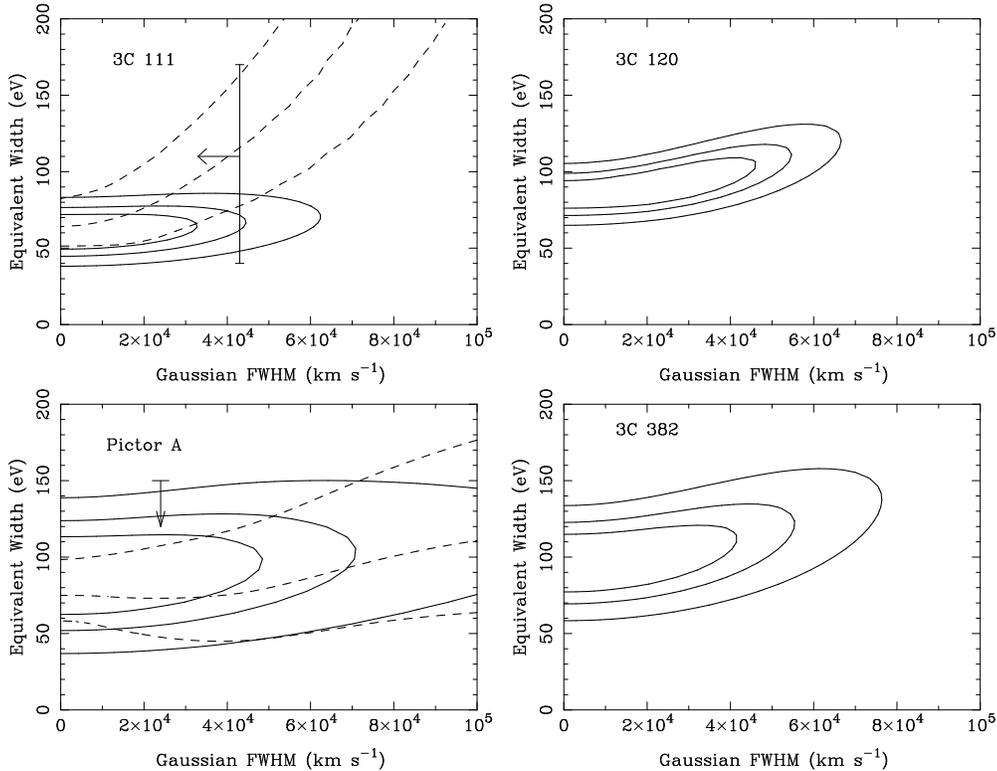,width=6.5in,rheight=4.8in,angle=-90}}
\caption{\footnotesize
68\%, 90\%, and 99\% confidence contours in the $EW$--$FWHM$
plane for the Fe~K$\alpha$ lines. The dashed lines in the case of 3C~111
and Pictor~A are the 68\%, 90\%, and 99\% upper limits obtained from {\it
ASCA} spectra by Eracleous \& Halpern (1998a,b). The error bar in the
3C~111 plot is the measurement of the line by Reynolds et al. (1998) using
{\it ASCA}, while the upper limit in Pictor~A comes from the {\it SAX}
spectrum of Padovani et al. (1999).}
\end{figure}

\subsection{The Fe~K$\alpha$ Line}

To study the properties of the Fe~K$\alpha$ line we modeled its
profile as a Gaussian of energy dispersion, $\sigma$, and intensity,
$I_{\rm Fe\,K\alpha}$. Because of the low energy resolution of the PCA,
more sophisticated models for the line profile are not warranted. For
each object we used the best-fitting continuum model and we ignored the
HEXTE spectra since they do not extend to energies below 20~keV. After
checking that the line centroid energies were consistent with the value
for ``cold'' Fe, we fixed them to the nominal value based on the
redshift of each object. We scanned the $I_{\rm Fe\,K\alpha}$--$\sigma$
plane for regions where the fit was acceptable, allowing the continuum
parameters to vary freely. We found that variations in the continuum
parameters were negligible throughout this plane, which we attribute to
the fact that the continuum is well constrained over a wide range of
energies. As a result the intensity of the line can be converted to an
equivalent width ($EW$) via a simple scaling. We, therefore, present
the outcome of the parameter search in the form of confidence contours
in the $EW$--$FWHM$\footnote{$FWHM$ is the full {\it velocity} width of
the line at half maximum.} plane, which we display in Figure~7. These
results are also summarized in Table~3. The lines are unresolved in all
cases, with upper limits on the $FWHM$ ranging from 44,000 to
70,000~km~s$^{-1}$ (at 90\% confidence). The $EW$s range from 60 to
90~eV with uncertainties around 30--50\% (at 90\% confidence). 

In the case of 3C~111 and Pictor~A, the results obtained here are
consistent with the results of {\it ASCA} and {\it SAX} observations in
which the Fe~K$\alpha$ line was either marginally detected (3C~111;
Reynolds et al. 1998; Eracleous \& Halpern 1998b) or not detected at
all (Pictor~A; Eracleous \& Halpern 1998a; Padovani et al. 1999). In
Figure~7 we overlay the {\it ASCA} upper limits on the {\it RXTE}
confidence contours for reference. In the case of 3C~120 and 3C~382
there is a significant discrepancy between the parameters determined
from the {\it ASCA} spectra and those we determine here: the lines
appear to be much broader and much stronger in the {\it ASCA} spectra
than in the {\it RXTE} spectra. In particular, Grandi et al. (1997)
find that the Fe~K$\alpha$ line of 3C~120 has $FWHM=91,000~{\rm
km~s^{-1}}$ and $EW=380$~eV while Reynolds (1997) finds that the
Fe~K$\alpha$ line of 3C~382 has $FWHM=197,000~{\rm km~s^{-1}}$ and
$EW=950$~eV. These velocity widths seem uncomfortably large, as pointed
out by the authors themselves. In view of the jet inclination angles
the line widths imply that some of the line-emitting gas moves faster
than light.  We suspect that the discrepancy is due to incorrect
determination of the continuum in the {\it ASCA} spectra (see Wo\'zniak et
al. 1998 and the discussion by Sambruna et al. 1999). This may be related
to the fact that the sensitivity of the  {\it ASCA} SIS, the instrument
most commonly used to measure the lines, is extremely low at $E>8$~keV.
Another possible cause for this discrepancy is calibration uncertainties
in the {\it ASCA} SIS below 1~keV, which affect the detection of a soft
excess, hence the determination of the spectral index. In fact, Wo\'zniak 
et al. (1998) find that if the continuum in the {\it ASCA} spectra of
3C~120 and 3C~382 is modeled as a broken power law (to account for a
soft soft excess) the measured $FWHM$ and $EW$ of the Fe~K$\alpha$ line 
become considerably smaller and consistent with our findings. It also
noteworthy, however, that Wo\'zniak et al. (1998) have almost certainly
underestimated the $EW$ of the Fe~K$\alpha$ line of 3C~120 since they
assumed the line to be unresolved. We have also been investigating the
cause of this discrepancy through simultaneous {\it ASCA} and {\it RXTE}
observations of  3C~382, the results of which we will report in a
forthcoming paper. 

\section{Discussion}

\begin{figure}[t]
\centerline{\psfig{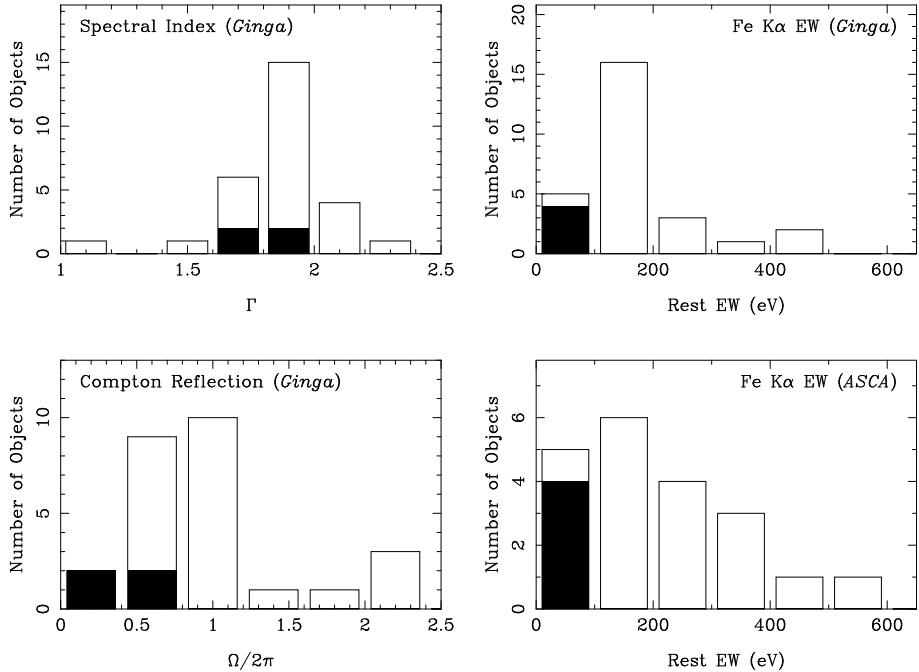}}
\caption{\footnotesize
The distribution of spectral parameters of Seyfert~1 galaxies
observed by {\it Ginga} (Nandra \& Pounds 1994) with the BLRGs observed
by {\it RXTE} marked for comparison. The distribution of $EW$s of Seyfert~1
galaxies observed {\it ASCA} (Nandra et al. 1997b) is also shown. Seyfert~1
galaxies are represented by hollow bins and BLRGs are represented by filled
bins.}
\end{figure}

\subsection{Comparison With Seyfert 1 Galaxies}

To search for systematic differences between BLRGs and  Seyfert~1
galaxies, we compare the X-ray spectral properties of the two classes
of objects using the results of our observations. In particular we
compare the photon indices, the strength of the Compton reflection
hump (parameterized by $\Omega/2\pi$), and the $EW$ of the Fe~K$\alpha$
line. We use the collection of Seyfert galaxies observed with {\it
Ginga} (Nandra \& Pounds 1994) as a comparison sample since the {\it
Ginga} LAC has a broad enough bandpass to allow a measurement of the
Compton reflection hump. We also use the collection of Seyfert galaxies
observed by {\it ASCA} (Nandra et al. 1997b) as an additional comparison
sample for the $EW$ of the Fe~K$\alpha$ line since the uncertainties in
the determination of the $EW$ from the {\it Ginga} spectra are relatively
large.  In Figure~8 we show the distribution of the above parameters
(spectral index, $\Omega/2\pi$, and Fe~K$\alpha$ $EW$) among Seyfert~1
galaxies with the values for the 4 BLRGs observed by {\it RXTE} superposed
as filled bins for comparison. A visual inspection of the histograms shows
that the spectral indices of BLRGs and Seyfert~1s are fairly similar
but the Compton reflection humps and Fe~K$\alpha$ lines of BLRGs are
considerably weaker than those of Seyfert~1s. Application of the
Kolmogorov-Smirnov (KS) test shows that the distributions of
Compton-reflection strengths and Fe~K$\alpha$ $EW$s are indeed
significantly different between Seyferts and BLRGs (chance
probabilities of 2\% for the former and 4\% for the latter; if the {\it
ASCA} $EW$ measurements are used instead of the {\it Ginga}
measurements, the chance probability is 3\%). In the case of the
spectral indices the KS test gives a 41\% probability that the spectral
indices of Seyferts and BLRGs were drawn from the same parent
population. Sambruna et al. (1999) reach a similar conclusion based
on the larger sample of BLRGs observed by {\it ASCA}.
To investigate whether the weakness of the Fe~K$\alpha$
lines of BLRGs is  a consequence of the X-ray Baldwin effect (Nandra et
al. 1997c) we also plot the BLRGs observed by {\it RXTE} in the
$EW$-$L_{\rm X}$ diagram shown in Figure~9. This figure shows clearly
that the BLRGs have weaker lines than Seyfert~1s of comparable X-ray
luminosity, supporting the outcome of the KS test presented above.
These results are in agreement with the conclusions of Wo\'zniak et al.
(1998) and Sambruna et al. (1999), who found that radio-loud AGNs in
general, and BLRGs in particular have weaker Fe~K$\alpha$ lines and
Compton reflection humps than their radio-quiet counterparts. They are
also supported by the results of {\it SAX} observations by Padovani et
al. (1999) and Grandi et al. (1998,1999). This raises the question of
the origin of the observed differences, which we discuss below.

\begin{figure}[t]
\centerline{\psfig{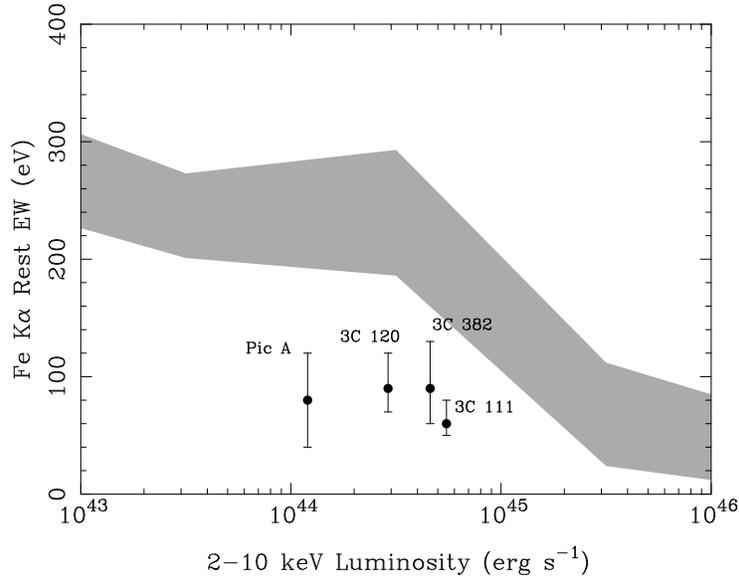}}
\caption{\footnotesize
The location of the BLRGs observed by {\it RXTE} in the
$EW$--$L_{\rm X}$ plane. The shaded band shows the region occupied by
Seyfert~1 galaxies in this plane, according to Nandra et al. (1997c); the
width of the band indicates the r.m.s. dispersion above and below the mean
$EW$ in that luminosity bin.}
\end{figure}

\subsection{Interpretation}

Possible causes of the systematic differences between radio-loud and
radio-quiet AGNs were discussed by Wo\'zniak et al. (1998) and by
Eracleous \& Halpern (1998a). Wo\'zniak et al. (1998) favored a
scenario in which the Fe~K$\alpha$ is produced in the obscuring torus
invoked in the unification schemes for type~1 and type~2 AGNs. A
column density in the reprocessing medium of $N_{\rm H}\sim
10^{23}~{\rm cm^{-2}}$ was found to reproduce the equivalent width of
the Fe~K$\alpha$ line. In this picture the reflection hump is still
due to Compton scattering in the accretion disk. However the continuum
is postulated to to be beamed away from the disk so that the
illumination of the disk is ineffective with the result that Compton
reflection and line emission from the disk made a relatively small
contribution to the observed spectrum of BLRGs. Eracleous \& Halpern
(1998a) concluded that there were two equally viable explanations for
the weakness of the Fe~K$\alpha$ line in the {\it ASCA} spectra of
Pictor~A and 3C~111: either the iron abundance in BLRGs is low (it
need only be lower by a factor of 2 compared to Seyferts), or the
solid angle subtended by the reprocessor to the primary X-ray source
is about a factor of 2 smaller in BLRGs than in Seyferts. Another
possibility worth examining is that the disks of BLRGs are in a higher
ionization state than those of Seyfert galaxies.  In such a case,
photons emerging from the disk will be Compton scattered by electrons
in the ionized skin of the disk, with the result that both the
Fe~K$\alpha$ line and the Fe~K edge are smeared out (Ross, Fabian, \&
Brandt 1996; Ross, Fabian, \& Young 1999) and appear weaker.  Finally,
we mention one more possible explanation: X-rays with a featureless
spectrum from the jets of BLRGs ``dilute'' the X-ray spectrum from the
central engine, which is otherwise similar to that of Seyferts. This
possibility was dismissed by both Wo\'zniak et al.  (1998) and by
Eracleous \& Halpern (1998a), but we re-examine it here in the light
of the latest observational results. We evaluate the merits and
applicability of these scenarios below.

\begin{enumerate}
\rightskip 0pt \pretolerance=100

\item {\it Continuum beamed away from the accretion disk resulting in weak
Fe~K$\alpha$ emission and Compton reflection from it; additional
Fe~K$\alpha$ emission from obscuring torus:} This explanation  requires
{\it mild} beaming by a sub-relativistic outflow, so that the X-ray beam
has an opening angle that is just right to illuminate the obscuring torus
effectively, while it it illuminates the accretion disk only ``mildly'' so
that the strength of the Compton reflection hump from the disk is
suppressed. This scenario suffers from the following drawbacks,
which make it untenable:

\begin{enumerate}
\item
It is contrary to observational evidence showing that the jets of BLRGs are
highly relativistic.  In particular, such jets are thought to have bulk
Lorentz factors of $\gamma=(1-\beta^2)^{-1/2} \gtrsim 10$ (Padovani \& Urry
1992; Ghisellini et al. 1993, 1998), which implies that their emission is
tightly focused in a cone of opening angle $\psi\approx 1/\gamma \lesssim
6^{\circ}$. This means that neither the accretion disk nor the obscuring
torus can be illuminated by the primary X-ray source, if this is
associated with a jet. Postulating that the X-ray continuum is beamed away
from the disk without being associated with the jet is an {\it ad hoc}
solution with no clear physical foundation. In fact, Reynolds \& Fabian
(1997) consider beaming of the X-ray continuum produced in a disk corona
and conclude that the beamed emission is directed in large part towards 
the disk. The result is that the $EW$ of the Fe~K$\alpha$ line is {\it
enhanced}.
\item
If the accretion disk is responsible for producing a Compton reflection
hump, then it should also be producing Fe~K$\alpha$ emission, which should
be added to the emission from the obscuring torus. If this were the case,
the Fe~K$\alpha$ $EW$ should have been disproportionately large compared to
the strength of the Compton reflection hump, since it would include
contributions from two different sources. The observations, however, show
that both the strength of the Compton reflection hump and the $EW$ of the
Fe~K$\alpha$ line are weaker by about the same factor compared to Seyfert
galaxies.
\item
Part of the motivation of Wo\'zniak et al. (1998) for associating
the source of the Fe~K$\alpha$ line with the obscuring torus was their
claimed lack of correlated variations of the continuum and line
intensities in 3C~390.3. This conclusion is not justified by their data,
however, because of the large uncertainties in the line intensity: a
factor of 2 at 68\% confidence (see their Figure~7). Since the continuum
variations span a range of a factor of 2, it is next to impossible to find
correlated variations in the line intensities when the error bars are as
large as they are. 
\end{enumerate}

Independently of the above arguments, this interpretation can be tested
observationally using the profiles of the Fe~K$\alpha$ lines. If the lines
are produced in the accretion disk, their profiles should be considerably
broader than if they are produced in the obscuring torus, with very
distinct asymmetries caused by Doppler and gravitational redshift.

\item {\it Low Iron Abundance:} A low iron abundance would have a very
clear observational consequence: it would make the Fe~K$\alpha$ lines
weaker but at the same time it would make the Compton reflection humps
stronger by reducing the opacity above the Fe~K edge (George \& Fabian
1991; Reynolds, Fabian, \& Inoue 1995). The {\it RXTE} spectra
presented here contradict this interpretation since both the
Fe~K$\alpha$ lines and the Compton reflection humps are weak. We have
specifically tested the hypothesis that the iron abundance is low with
negative results (see \S4.1 and Figure~6). Based on these findings we
conclude that a low iron abundance is not a viable interpretation. 

\item {\it Small solid angle subtended by the reprocessor to the
primary X-ray source:} The weakness of both the Fe~K$\alpha$ line and
the Compton reflection hump can be explained if the reprocessing medium
subtends a relatively small solid angle to the primary X-ray source. In
the case of Seyfert galaxies, the primary X-ray source is thought to
make up a hot corona overlaying the disk proper, which serves as the
reprocessor. In such a picture the reprocessor covers half the sky as
seen from the primary X-ray source, i.e., the solid angle is
$\Omega/2\pi=1$, in reasonable agreement with the observed X-ray
spectra of Seyfert galaxies. In the case of BLRGs, on the other hand,
the {\it RXTE} spectra indicate that $\Omega/2\pi \lesssim 0.5$. This
can be explained in a picture where the primary X-ray source is a
quasi-spherical ion torus (or ADAF) occupying the inner disk and the
reprocessor is the geometrically thin, optically thick outer disk. The
solid angle subtended by the outer disk to the ion torus is
$\Omega/2\pi < 0.5$ on  geometrical grounds (Chen \& Halpern 1989;
Zdziarski, Lubi\'nski \& Smith 1999), in agreement with our findings.
This is our preferred interpretation over all the others we consider
here and has the following additional attractive features: (a) the
association of ion tori with BLRGs is appealing because these
structures offer a way of producing radio jets as we have mentioned in
\S1, and (b) such an accretion disk structure also explains the
double-peaked profiles of the Balmer lines in some of these objects
(Pictor~A, 3C~382, and 3C~390.3; Chen \& Halpern 1989; Eracleous \&
Halpern 1994; Halpern \& Eracleous 1994). The profiles of the
Fe~K$\alpha$ lines offer a way of testing this scenario further: if the
lines do indeed originate in the outer parts of the accretion disk at
radii $R\gtrsim 100~R_{\rm g}$ (where $R_{\rm g}\equiv GM/c^2$ is the
gravitational radius, with $M$ the mass of the black hole), then their
profiles should be narrower than those observed in Seyfert galaxies,
although still skewed, asymmetric, and possibly double-peaked.

\item{\it Smearing by Compton scattering in an ionized accretion
disk:} If the disk is photoionized so that the ionization parameter
reaches values of $\xi > 10^4$, photons emerging from the disk will be
Compton scattered by hot electrons in the disk atmosphere.  This
process will smear out of the Fe~K$\alpha$ line and the Compton
reflection and make them appear weaker (see Ross et al. 1996, 1999).
We consider this an unlikely explanation for our results because the
observed spectra lack the distinct signature of this process.  Namely,
at $\xi > 10^3$, the Fe~K$\alpha$ lines is considerably broadened by
Compton scattering while at $\xi > 10^4$, the line centroid shifts to
energies greater than 6.7~keV because most of the iron atoms are
highly ionized. Neither of these effects is observed in either the
{\it RXTE}~PCA spectra presented here or in published {\it ASCA}
spectra.

\item {\it Dilution of the X-ray spectrum of the central engine by
beamed emission from the jet:} At first glance this is a plausible
interpretation since a flux from the jet comparable to the flux from
the central engine would provide the necessary dilution. 
This hypothesis does not withstand close scrutiny, however, for the
following reasons: (a) as shown by Wo\'zniak et al. (1998) the spectra of
BLRGs {\it cannot} be described as the sum of two power-law components,
which should have been possible if they included comparable contributions
from the jet and the central engine, (b) the light curves of our targets
(\S3 and Figure~2) do not show a great deal of variability on time scales
up to a few days, contrary to what is observed in blazars; 3C~120 shows
the largest variability amplitude, consistent with the small inclination
angle of the jet to the  line of sight (Table~1), which make the
contribution of the jet to the  observed X-ray emission rather
significant, (c) the X-ray spectral indices of BLRGs are systematically
different from those of blazars, which are typically around 1.4 to 1.5
(Kubo et al. 1998), and (d) 3C~120 is the object where  dilution of the
observed spectrum by emission from the jet is expected to be most severe
because of the very small inclination angle; nevertheless both an
Fe~K$\alpha$ line and a Compton reflection hump are detected, when they
should have been undetectable.

\end{enumerate}

\section{Conclusions and Future Prospects}

Our study of the hard X-ray spectra of 4 BLRGs observed with {\it RXTE}
has shown them to be systematically different from those of Seyfert
galaxies. In particular, the Fe~K$\alpha$ lines and Compton reflection
humps in the spectra of BLRGs are at least a factor of 2 weaker than those
observed in the spectra of Seyfert galaxies. This result is consistent
with the conclusions of previous studies and is supported by the results
of {\it SAX} observations of  the same targets. After examining several
possible explanations for this difference, we conclude that the most
likely one is that the solid angle subtended by the source of these
spectral features to the primary X-ray source is a factor of 2 smaller in
BLRGs than in Seyferts. Since this reprocessing medium is thought to be
the accretion disk, we interpret this difference as the result of a
difference in the accretion disk structure between BLRGs and Seyferts.
More specifically, we argue that if the inner accretion disks of BLRGs
have the form of an ion-supported torus (or an ADAF) that irradiates the
outer disk, then the observed differences can be explained. We find this
explanation particularly appealing because ADAFs offer a possible way of
producing the radio jets in these objects and because such a disk
structure can also account for the double-peaked Balmer line profiles
observed in these objects. 

This interpretation, as well as some of the other scenarios we have
considered, can be tested further by studying the profiles of of the
Fe~K$\alpha$ lines. Unfortunately, such a test has not been possible so
far using spectra from {\it ASCA}, because of their low signal-to-noise
ratio. It will be possible, however, to carry out the test using spectra
from upcoming observatories such as {\it XMM} and {\it Astro-E}.
Correlated variations of the intensity of the Fe~K$\alpha$ line and the
strength of the X-ray continuum afford an additional observational test of
these ideas. The light travel time between the ion torus and the outer
accretion disk is  $\tau_{\ell} \approx 1.7\,(R/300\,R_{\rm
g})\,(M/10^8\,M_{\odot})$~days, which means that one may expect lags of
this order between line and continuum variations. If, on the other hand,
the line is produced very close to the center of the disk, the lags should
be considerably shorter, while if the line is produced in the obscuring
torus, the lags should be on the order of several years. Monitoring
campaigns with {\it RXTE} may provide the data needed for such a test.  

\acknowledgements 

We are grateful to Karen Leighly and Niel Brandt for useful discussions 
and we thank the anonymous referee for thoughtful comments. This work was
supported by NASA through grants NAG5--7733 and NAG5--8369. During the
early stages of this project M.E. was based at the University of
California, Berkeley, and was supported by a Hubble Fellowship (grant
HF-01068.01-94A from the Space Telescope Science Institute, which is
operated for NASA by the Association of Universities for Research in
Astronomy, Inc. under contract NAS~5--2655). R.M.S. also acknowledges
support from NASA contract NAS--38252.

\end{document}